\documentclass[preprint]{aastex}

\shorttitle{MIR-EXCESS OF EARLY-TYPE GALAXIES}
\shortauthors{Ko et al.}

\begin{document}


\title{STELLAR POPULATIONS OF EARLY-TYPE GALAXIES WITH 
MID-INFRARED EXCESS EMISSION}

\author{Jongwan Ko\altaffilmark{1,2}, Haeun Chung\altaffilmark{3,4}, Ho Seong Hwang\altaffilmark{3}, Jong Chul Lee\altaffilmark{1}}
\altaffiltext{1}{Korea Astronomy and Space Science Institute, Daejeon 34055, Korea}
\altaffiltext{2}{University of Science and Technology, Daejeon 305-350, Korea}
\altaffiltext{3}{School of Physics, Korea Institute for Advanced Study, 85 Hoegiro, Dongdaemun-gu, Seoul 02455, Korea}
\altaffiltext{4}{Astronomy Program, Department of Physics and Astronomy, Seoul National University, Seoul, Korea}

\email{jwko@kasi.re.kr}

\begin{abstract}

We present a stellar population analysis of quiescent 
 (without H$\alpha$ emission) and bright ($M_{r}$ $<$ $-$21.5) 
 early-type galaxies (ETGs) with recent star formation.
The ETGs are selected from a spectroscopic sample of SDSS galaxies at 
 0.04 $<$ $z$ $<$ 0.11 with {\it WISE} mid-infrared (IR) and {\it GALEX} 
 near-ultraviolet (UV) emissions.  
We stack the optical spectra of ETGs with different amounts of mid-IR
 and near-UV excess emissions to measure  
 the strength of 4000 \AA{} break $D_{n}$4000 and the width of Balmer 
 absorption line H$\delta_{A}$ that are indicative of 
 recent ($\sim$1 Gyr) star formation activity.
The {\it WISE} [3.4]$-$[12] colors show stronger correlations with the 
 spectral features than NUV$-r$ colors. 
We fit to the stacked spectra with a spectral fitting code, 
 STARLIGHT, and find that the mass fraction of young ($\leq$1 Gyr) and 
 intermediate-age ($\sim$1$-$5 Gyr) stars in the ETGs with mid-IR excess 
 emission is $\sim$4$-$11\%, depending on the template spectrum used for the 
 fit.
These results show that the ETGs with mid-IR excess emission have 
 experienced star formation within the last 1$-$5 Gyr and that 
 the mid-IR emission is a useful diagnostic tool for probing 
 recent star formation activity in ETGs.
\end{abstract}


\section{INTRODUCTION}

The mid-infrared (IR) data for nearby early-type galaxies (ETGs) become 
 increasingly important because many observational data of ETGs show a 
 clear excess over stellar photospheric emission
 at 8$-$24 $\mu$m (e.g., Knapp et al. 1992; Athey et al. 2002;
 Bressan et al. 2006; Ko et al. 2009, 2012, 2013; Shim et al. 2011; 
 Hwang et al. 2012; Martini et al. 2013; Rampazzo et al. 2013).
In old stellar systems (i.e., ETGs), the most likely origin of 
 such mid-IR excess is circumstellar dust around asymptotic giant branch 
 (AGB) stars (e.g., Bressan et al. 1998; Piovan et al. 2003; 
 Villaume et al. 2015).
For example, Villaume et al. (2015) showed that $Spitzer$ 
 mid-IR photometric and spectroscopic data for ETGs 
 (from Martini et al. 2013 and Rampazzo et al. 2013) are well explained 
 by their AGB dust model in the wavelength range 
 where there are no emission lines.
They also show that dust models without AGB dust cannot fit to the mid-IR 
 data for ETGs detected at 70 and 160 $\mu$m.
  

Although the detailed contribution of AGB dust to the mid-IR spectral 
 energy distribution (SED) of a galaxy is not well
 understood yet (e.g., Kelson \& Holden 2010; Melbourne \& Boyer 2013), 
 recent models consistently suggest that the mid-IR emission from AGB 
 dust could be a proxy for the existence of intermediate-age stars in 
 old stellar systems with little diffuse dust.
This is because the mid-IR SED is very sensitive to the 
 (even small) amount of stellar mass formed within $\sim$0.3$-$3 Gyr, 
 allowing us to trace recent star formation 
 (e.g., see Villaume et al. 2015 and references therein).  
However, to better understand the mid-IR excess emission of ETGs, it is 
 necessary to quantify the amount of intermediate-age stars in ETGs where 
 the mid-IR emission is attributed to AGB dust.
The most promising way is to investigate their stellar populations 
 derived from age-sensitive features in optical absorption line 
 spectroscopy (e.g., Temi et al. 2005; Bregman et al. 2006; 
 Conroy et al. 2014).

 
Ko et al. (2013; hereafter Ko13) studied the mid-IR properties of 
 local ETGs on the optical red sequence. 
We found that around half of 648 quiescent red-sequence galaxies have
 mid-IR excess emission over the stellar component.
If we consider only bright ($M_{r}$ $<$ $-$21.5) ETGs,
 the fraction of mid-IR excess is still 39\%.
We also found that the ETGs with mid-IR excess 
 have the strength of the 4000 \AA{} break slightly smaller than those 
 without mid-IR excess (i.e., existence of more young stars).
However, we did not find any significant difference in Balmer 
 absorption line index H$\delta_{A}$ between the two because the 
 signal-to-noise ratio (S/N) of each individual SDSS spectrum is not high enough 
 to examine the difference.

Here we use the ETG sample of Ko13 to quantify the amount of 
 intermediate-age stars; we perform a stellar population analysis of 
 the stacked SDSS spectra of these ETGs.
We focus on the mid-IR excess of ETGs and quantify the amount of 
 intermediate-age stars through a stellar population analysis 
 of the SDSS stacked spectra.
If the mid-IR emission of ETGs is produced by intermediate-age stars 
 formed within the last 1$-$5 Gyr, the absorption features 
 in the optical spectra can provide important information 
 on the age of their stars.
Section 2 describes the observational data we use. 
In Section 3 we examine spectral features of the ETGs with mid-IR 
 excess emission, and conclude in Section 4. 
Throughout, we use the AB magnitude system, and 
 adopt flat $\Lambda$CDM cosmological parameters: 
 $H_0 = 70$ km s$^{-1}$ Mpc$^{-1}$, $\Omega_{\Lambda}=0.7$ and 
 $\Omega_{m}=0.3$.

\section{DATA AND SAMPLE}

We use the observational data and the ETG sample constructed in Ko13. 
Here we give a brief summary of the data and galaxy sample.

\subsection{Sample: Quiescent ETGs on the Red-sequence}

In a spectroscopic sample of galaxies in the SDSS data release 7 
 (DR7; Abazajian et al. 2009), we first identified ETGs using the Korea 
 Institute for Advanced Study (KIAS) value-added galaxy catalog that 
 provides galaxy morphological information (Choi et al. 2010). 
To identify optically quiescent ETGs, we plot these ETGs in the 
 color-magnitude diagram (CMD), and select the galaxies in the 
 red sequence (see Figure 4 in Ko13). 
We then removed the galaxies with H$\alpha$ equivalent width 
 $<$ $-$1 \AA{} to select only ``quiescent'' galaxies.
We further removed the galaxies with active galactic nuclei (AGNs) 
 because they can produce mid-IR emission.
Most AGNs are removed from the H$\alpha$ equivalent width cut.
The remaining AGNs are removed using the spectral types determined by 
 the criteria of Kewley et al. (2006). 
We also used the {\it WISE} color-color selection criteria of 
 Jarrett et al. (2011) to remove dusty AGNs.  
We also excluded the highly inclined ($i$-band isophotal axis ratio $b/a$ 
 less than 0.6; see Choi et al. 2007) galaxies to reject dust-reddened 
 star-forming galaxies. 
In the result, we obtain a sample of 397 bright ($M_{r}$ $<$ $-$21.5), 
 quiescent (without H$\alpha$ emission), red ETGs without AGNs
 (see Section 2.2 in Ko13 for details).
We compile the multi wavelength data for these galaxies including 
 {\it WISE} mid-IR and {\it GALEX} near-UV for the following analysis 
 (Hwang et al. 2010). 


\subsection{Subsamples: Classifying ETGs Using Near-UV and Mid-IR Excess Emissions}

\begin{figure}[h!]
\centering
\includegraphics[width=14cm]{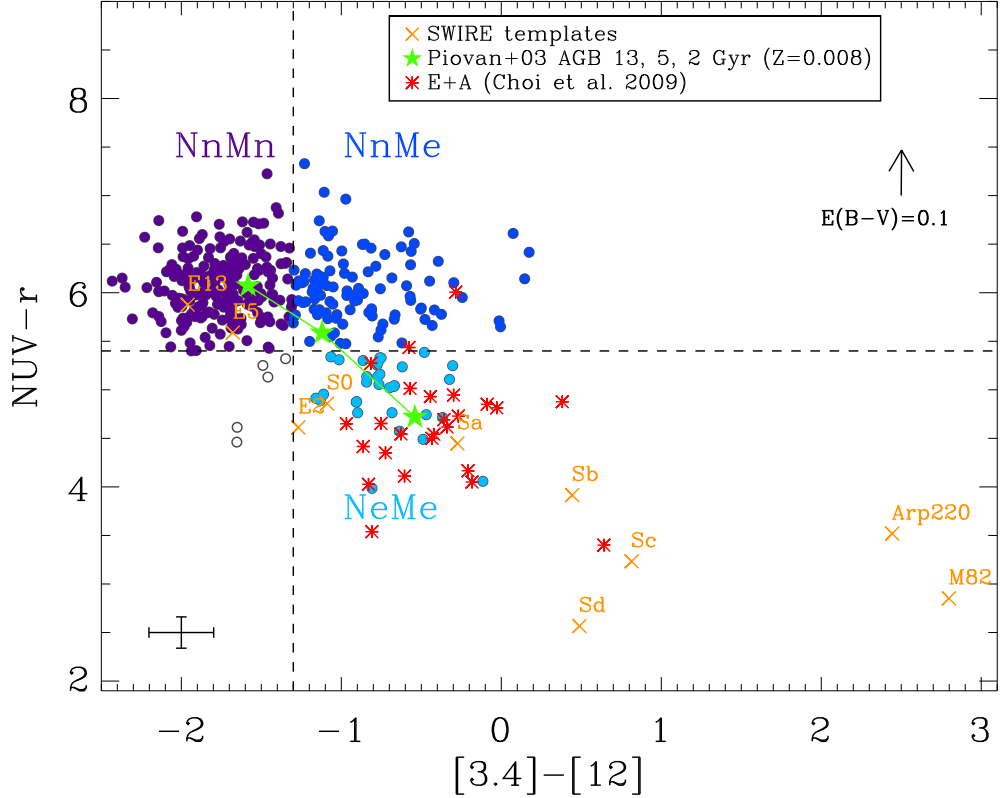}
\caption{[3.4]$-$[12] vs. NUV$-r$ colors of bright 
		 ($M_{r}$ $<$ $-$21.5), quiescent (without H$\alpha$ emission) 
		 ETGs.
         Circles with different colors indicate different subclasses. 
         Asterisks represent bright E+A 
         galaxies with early-type morphology from Choi et al. 
         (2009).
         For reference, the extinction-free colors of models
         (which have different prescription of AGB dust) are 
         overplotted:          
         the SWIRE templates of Polletta et al. (2007) including three 
         ellipticals (2, 5, and 13 Gyr), five spirals (S0, Sa, Sb, Sc, 
         Sd), and two starbursts (M82 and Arp220); and
         single stellar population templates of Piovan et al. (2003)
         along light-weighted mean stellar ages (2, 5, and 13 Gyr) for 
         40\% solar metallicity (Z=0.008). 
         The vertical and horizontal dashed lines indicate the mid-IR 
         color threshold of [3.4]$-$[12] = $-$1.3 and the near-UV color 
         threshold of NUV$-r$ = 5.4, respectively. 
         The arrow shows the mean value of a reddening vector of 
         E(B-V) = 0.1 for the SWIRE templates, 
         and the cross in the the lower left corner 
         indicates median color errors.}         
\end{figure}

Figure 1 shows the distribution of [3.4]$-$[12] and NUV$-r$ colors 
 for 397 bright ETGs.
They show a wide range of colors in the near-UV and mid-IR.
We classify these ETGs into three subclasses based on NUV$-r$ and  
 [3.4]$-$[12] colors.
We adopt the cut of [3.4]$-$[12] = $-$1.3 (Ko13); 
 galaxies redder than this cut are considered to be relatively young 
 ($\lesssim$ 5 Gyr) in the mid-IR weighted mean stellar ages.
Because we have selected the galaxies without H$\alpha$ emission,
 AGNs, and highly inclined disks, the mid-IR emission is dominated by
 the circumstellar dust around AGB stars.
The NUV$-r$ cut is NUV$-r$ = 5.4 (Schawinski et al. 2007); 
 galaxies bluer than this threshold are considered to have  
  $\gtrsim$1\% of young ($\lesssim$ 1 Gyr) populations
 (see Fig. 6 in Ko13).
The three subclasses can be summarised as follows.

$\bullet$ ETGs with near-UV excess and mid-IR excess 
 (`NeMe' in Fig. 1): NUV$-r$ $<$ 5.4 and [3.4]$-$[12] $>$ $-$1.3.
These galaxies have both mid-IR and near-UV excess emissions.

$\bullet$ ETGs without near-UV excess but with mid-IR excess
 (`NnMe' in Fig.1): NUV$-r$ $>$ 5.4 and [3.4]$-$[12] $>$ $-$1.3.
These galaxies have only mid-IR excess emission over the stellar 
 component.
There are five ETGs with near-UV excess and without mid-IR excess, 
 but we do not analyze them because of small number statistics. 
Three of them are close to the boundaries, suggesting that they 
 may belong to other subclasses if we consider their errors in colors.

$\bullet$ ETGs without near-UV excess and mid-IR excess 
 (`NnMn' in Fig. 1): NUV$-r$ $>$ 5.4 and [3.4]$-$[12] $<$ $-$1.3.
These galaxies are completely quiescent in the optical, near-UV and 
 mid-IR.
Note that the galaxies with [3.4]$-$[12] $<$ $-$1.7 and NUV$-$r $>$ 5.7 
 are undetected at 12 $\mu$m and at NUV, respectively. 
Therefore, most galaxies in this subclass are indeed quiescent.

For comparison, we also show E+A galaxies from Choi et al. (2009) with
 similar absolute magnitudes ($M_{r}$ $<$ $-$21.5) and morphology 
 (i.e. ETGs) in Figure 1. 
E+As are post-starburst (within $\sim$1 Gyr) systems characterized by
 strong Balmer absorption lines and weak H$\alpha$ emission line 
 (e.g., Yang et al. 2008). 
Most E+A galaxies are located in the NeMe region
 (except two galaxies that have only mid-IR excess emission).
However, E+As have much bluer NUV$-r$ and redder [3.4]$-$[12] colors than
 NeMe galaxies.
Because E+As are not red-sequence galaxies, this means that they are 
 not in the same evolutionary stage as NeMe galaxies. 
In other words, E+As have much more young stars than NeMe galaxies, 
 thus E+As have mid-IR weighted mean stellar ages much younger than NeMe 
 galaxies.

\section{RESULTS AND DISCUSSION}

Here we use the SDSS optical spectra of the ETGs in the three subclasses 
 for a stellar population analysis. 
To increase the S/N of the spectra, we stack the galaxy spectra in each 
 subclass.

\subsection{SDSS Stacked Spectra of ETGs}

\begin{figure}[!h]
\centering
\includegraphics[width=14cm]{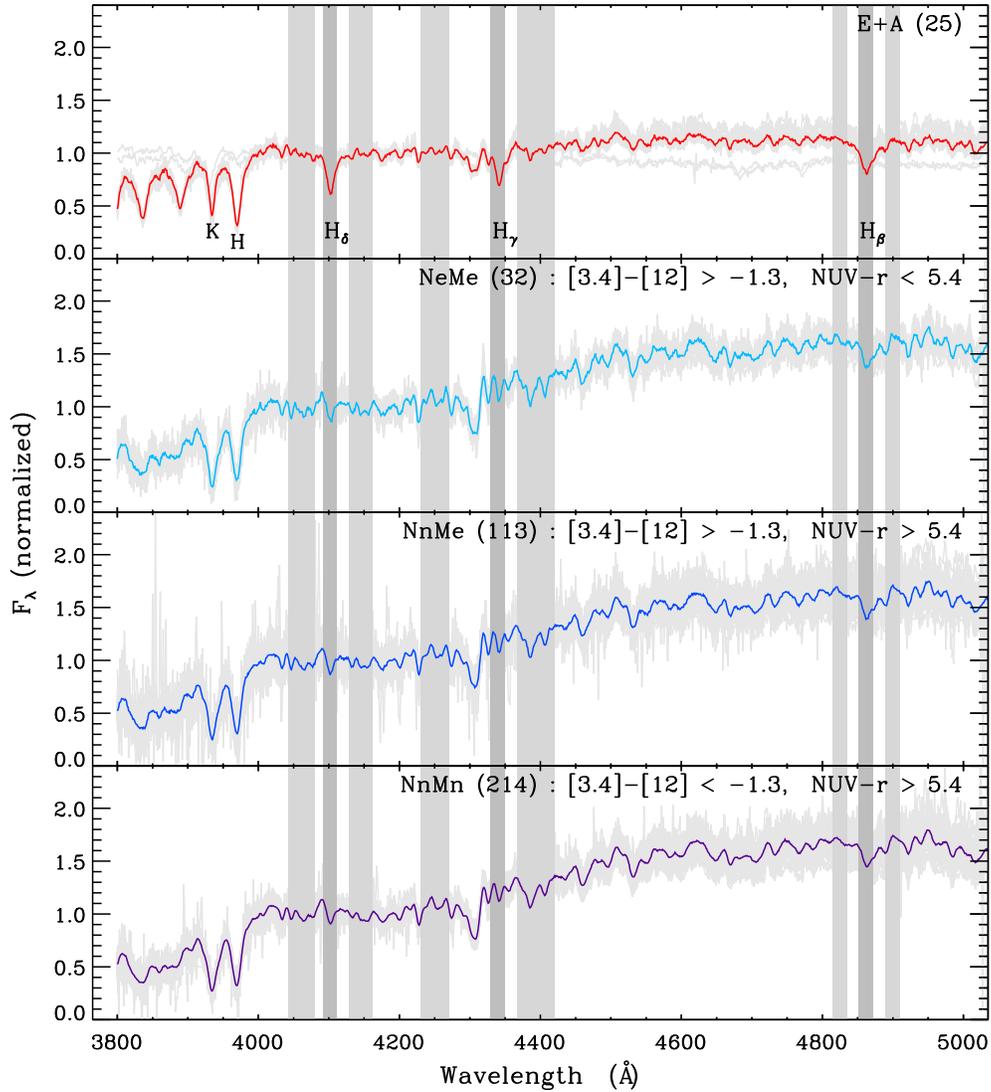}
\caption{SDSS stacked spectra of three subclasses of ETGs and 
	     of E+A galaxies.
	     The number in parenthesis in each panel is the number of 
	     galaxy spectra used for stacking.
         The light and dark gray-shaded regions show 
         the pseudo-continuum for the Balmer absorption line index and 
         the Balmer absorption line bandpass, respectively.         
         }         
\end{figure}

For each subclass of ETGs and E+As, we stack the SDSS spectra. 
We normalize the individual spectrum at rest-frame 4150$-$4250 \AA{} 
 and take the median for the stacking. 
The wavelength coverage for the stacked spectra is 3800$-$7650 \AA{}. 
The S/N of the stacked spectra is about 71-281 at 4092$-$4111 \AA{} 
 (around the H$\delta$ absorption passband), an order of magnitude 
 higher than the typical S/N of individual spectrum ($\sim$11).

We measure the strength of 4000 \AA{} break, $D_{n}$4000, and 
 Lick indices of Balmer absorption lines. 
We adopt the bandpass of $D_{n}$4000 defined by 
 Balogh et al. (1999), and the Lick indices defined by 
 Worthey \& Ottaviani (1997).
We adopt the continuum bandpasses of Moustakas \& Kennicutt (2006, 
 Table 9 therein) for Balmer absorption lines. 
Figure 2 shows the stacked spectra of three subclasses and E+As
 in the wavelength range 3700$-$5000 \AA{}.
The stacked spectra of three subclasses show that all of them
 are typical quiescent systems with strong 4000 \AA{} break,
 but the stacked spectrum of E+As shows different features 
 including a strong H$\delta$ absorption line.

The strength of 4000 \AA{} break correlates well with stellar ages, 
 and strong H$\delta$ absorption lines are attributed to 
 recent (0.1 $-$ 1 Gyr) bursts of star formation (Bruzual \& Charlot 
 2003; BC03).
Kauffmann et al. (2003) showed that the combination of $D_{n}$4000 
 and H$\delta_{A}$ is a good indicator of star formation 
 activity of galaxies over the past 1$-$2 Gyr with little 
 metallicity effect.
In Figure 3, we show the $D_{n}$4000 versus H$\delta_{A}$ for the
 BC03 simple stellar population (SSP) models with a solar metallicity and 
 the star formation history of SFR($t$) $\varpropto$ exp($-t$/$\tau$).
We use a formation time from 10 Gyr to 0.1 Gyr, and a prompt star 
 formation history of $\tau$ = 0.5 Gyr that is generally expected for 
 SDSS ETGs (e.g., Thomas et al. 2010; Li et al. 2015). 
E+As lie in a region, clearly different from the region of ETGs 
 where old populations are dominant with no significant recent bursts.
The ETGs with mid-IR excess (NeMe and NnMe) appear younger 
 than the ETGs without mid-IR excess (NnMn).
This is consistent with our previous results that the mid-IR flux is 
 sensitive to young and intermediate-age ($\geq$1 Gyr) populations.

Figure 3 also shows a hint of slightly younger ages of NeMe galaxies
 than NnMe galaxies, consistent with the difference in the mid-IR weight 
 mean stellar age in Figure 1. 
This means that galaxies with near-UV excess at a given mid-IR flux 
 have more young ($\leq$1 Gyr) stars than those
 without near-UV excess, 
 if the dust attenuation effects are negligible (see Figure 6 of Ko13).
This can suggest that NeMe galaxies evolve into NnMe galaxies within 
 $\sim$1 Gyr, consistent with previous studies; 
 Villaume et al. (2015) show that the mid-IR light of ETGs is 
 enhanced by a factor of 3 because of stars formed within
 $\sim$0.3$-$3 Gyr, while the near-UV light of ETGs remains 
 bright only for $\sim$1.5 Gyr (Yi et al. 2005).


\begin{figure}[!h]
\centering
\includegraphics[width=14cm]{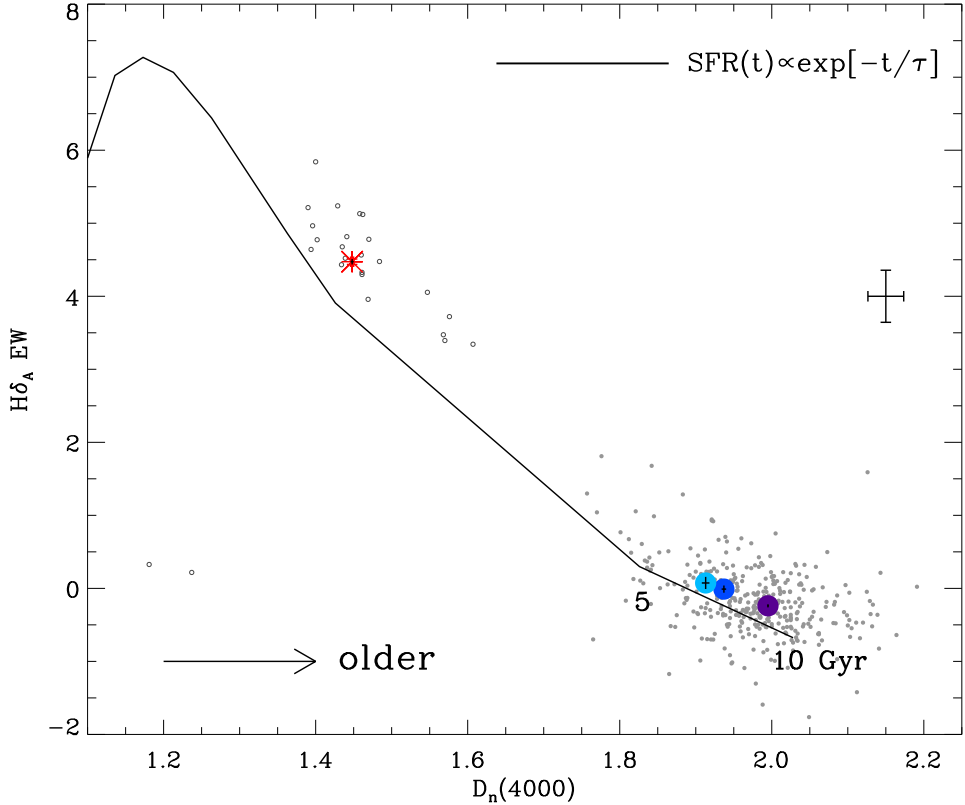}
\caption{H$\delta_{A}$ vs. $D_{n}$4000 for 
         the SDSS stacked spectra of three subclasses of ETGs and E+As
         (large symbols). 
         Colors are the same as in Figure 1.  
         Gray filled and open symbols represent individual spectra for 
         three subclasses and E+As, respectively.
         The large cross indicates median errors of individual spectrum.     
         The solid line represents a model of continuous star formation
         histories with a solar metallicity, a fixed $\tau = 0.5$ Gyr, 
         and different formation time (10 Gyr to 0.1 Gyr). 
         We indicate two formation ages of 5 and 10 Gyrs.}         
\end{figure}

\subsection{Tracing Recent Star Formation History of ETGs; via optical spectral features, near-UV and mid-IR excess emissions}

From a two-component SSP analysis of ETGs in Ko13, 
 we found that the small amount of young ($\leq$1 Gyr) populations 
 can make ETGs a significant change in [3.4]$-$[12] colors of ETGs, 
 similar to the case of NUV$-r$ colors (see Figure 6 of Ko13).
The SSPs with AGB dust further show that only the mid-IR data 
 can trace star formation longer than $\gtrsim$1 Gyr
 even though both near-UV and mid-IR are sensitive to star formation 
 over the past $\sim$1 Gyr.
 
To study correlations between optical spectral features and 
 near-UV/mid-IR data as recent star formation indicators, 
 we plot $D_{n}$4000, H$\delta_{A}$, and 
 $\langle$H$\beta\gamma\delta_{A}\rangle$ as a function of 
 [3.4]$-$[12] and NUV$-r$ colors of ETGs in Figure 4;  
 $\langle$H$\beta\gamma\delta_{A}\rangle$ is the average of three Lick 
 indices of Balmer absorption lines (H$\beta_{A}$, H$\gamma_{A}$ and H$
 \delta_{A}$). 
We measure the spectral features of both stacked spectra 
 (large color circles) and individual spectrum (gray circles). 
The Spearmann rank correlation test for the data from the individual 
 spectrum suggests that the correlation between the optical features and 
 [3.4]$-$[12] color is stronger than that between 
 the optical features and NUV$-r$ colors. 
Interestingly, [3.4]$-$[12] colors of NnMe galaxies differ from those of 
 other subclasses (i.e., NnMe galaxies show redder colors than NnMn 
 galaxies, but show bluer colors than NeMe galaxies), but NUV$-r$ colors of 
 NnMe galaxies are similar to those of NnMn galaxies. 
These results support that the mid-IR data probe the star formation 
 activity of ETGs within the last 1$-$2 Gyrs better than the near-UV 
 data. 
Therefore, many ETGs without young ($\lesssim$ 1 Gyr) stars (i.e., NnMe) 
 seem to have excess emission in the mid-IR because of their 
 intermediate-age ($\gtrsim$ 1 Gyr) stars or/and to have no near-UV 
 excess emission because of the small fraction of 
 young ($\lesssim$ 1 Gyr) stars if the dust attenuation effects are 
 negligible.


\begin{figure}[!h]
\centering
\plottwo{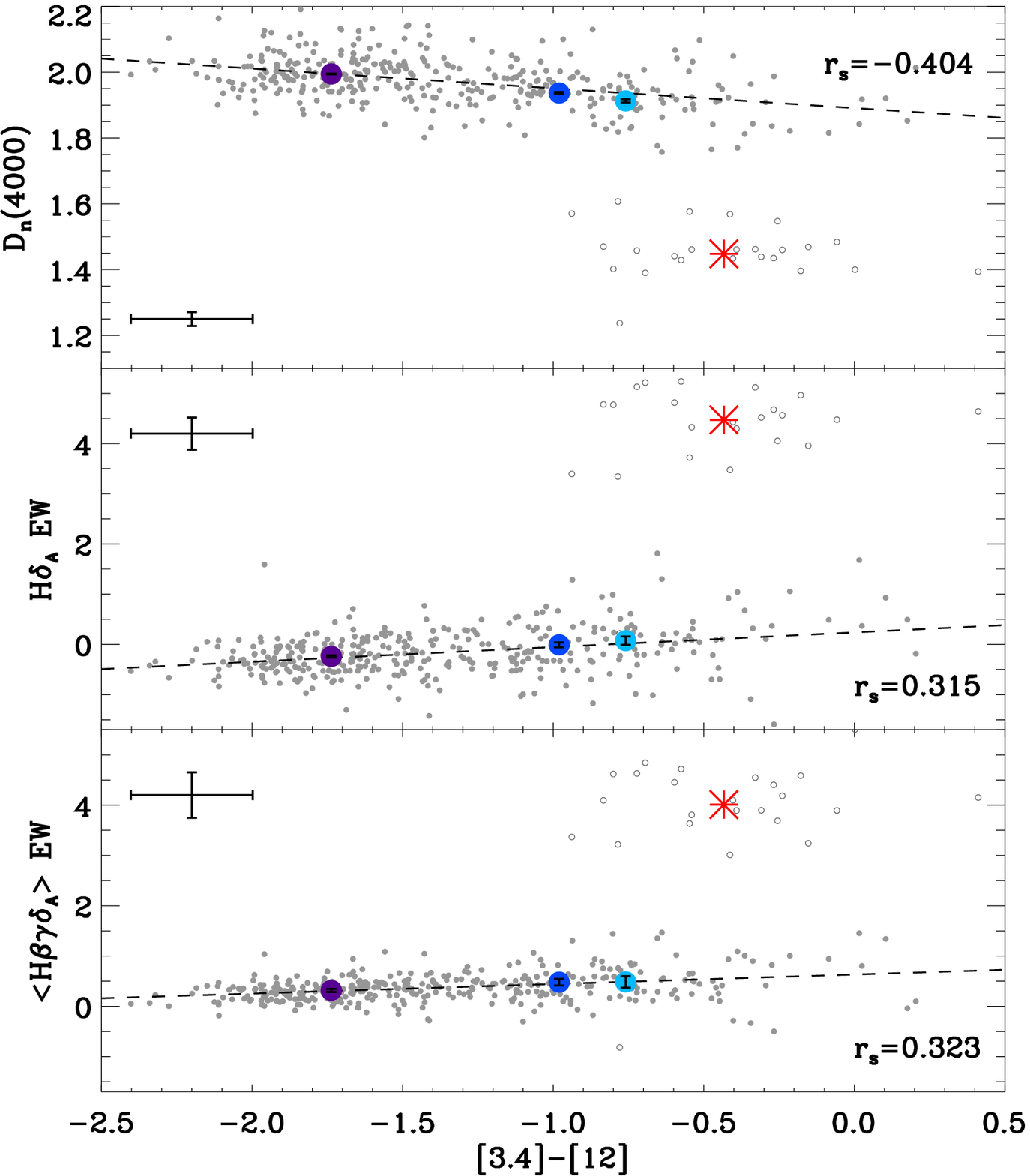}{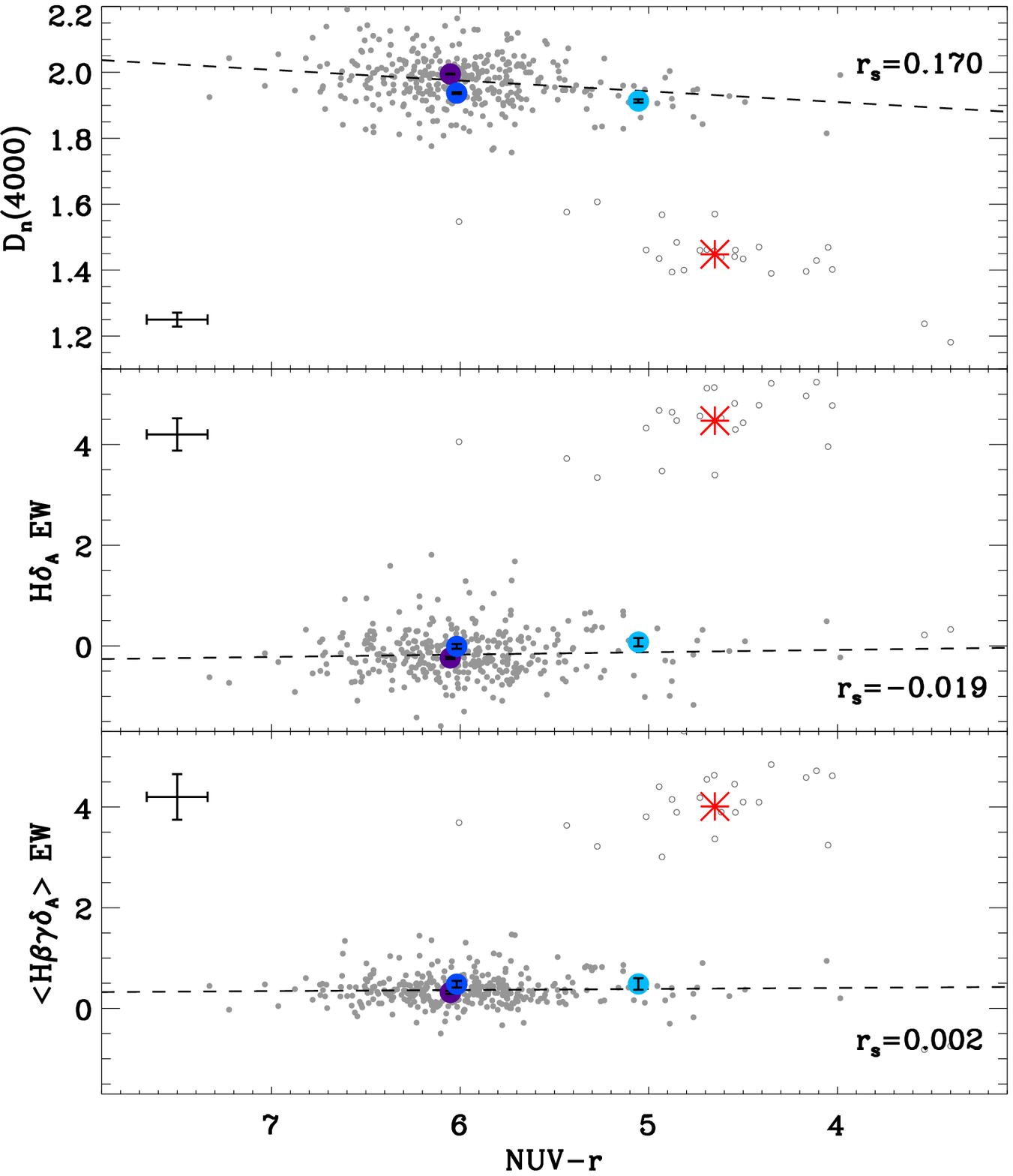}
\caption{4000 \AA{} break ($D_{n}$4000) and 
         Balmer absorption features (H$\delta_{A}$ and 
         $\langle$H$\beta\gamma\delta_{A}\rangle$, the average of the 
         equivalent widths of H$\beta_{A}$, H$\gamma_{A}$ and H$\delta_{A}$)
         as a function of [3.4]$-$[12] (\textit{Left}) 
         and NUV$-r$ (\textit{Right}) colors for 
         individual galaxies in the three subclasses (filled circles) 
         and E+A galaxies (open circles).
         The cross in the left corner of each panel indicates median 
         errors.
         The Spearman rank correlation coefficient $r_{s}$ between 
         colors and spectral features for the three subclasses of ETGs 
         are shown in the right corner of each panel.
         Large symbols represent the data from the stacked spectra of
         the ETGs in three subclasses and of E+As.
         }         
\end{figure}

\subsection{Stellar Populations of ETGs with Mid-IR Excess: Fit to the Optical Spectra with STARLIGHT}

For each stacked spectrum, we perform a decomposition of stellar 
 populations using the spectral fitting code STARLIGHT 
 (Cid Fernandes et al. 2005) in the wavelength range 
 3800$-$7650 \AA{}.
We first use 45 templates of BC03 model covering 15 ages 
 (between 1 Myr and 13 Gyr) and 3 metallicities (Z$=$0.004, 0.02, 
 0.05), generated from STELIB library (Le Borgne et al. 2003) with 
 Padova (1994) evolutionary tracks and Chabrier (2003) initial mass 
 function (IMF).
Figure 5 shows the result of the STARLIGHT fit to the stacked spectra 
 of the three subclasses and E+As.
We perform the fit with 100 different seeds for generating random 
 numbers (see the STARLIGHT manual for details), 
 and adopt the median of the results from the 100 fits.

As expected, all ETGs are dominated by a $\sim$10 Gyr stellar 
 population.
The mass fraction of young ($\leq$1 Gyr) and 
 intermediate-age ($\sim$1$-$5 Gyr) populations for the ETGs 
 with mid-IR excess (i.e., NnMe and NeMe) 
 is small ($\sim$4$-$5\%), but is not negligible compared to the 
 ETGs without mid-IR excess (NnMn).
E+As have young ($\sim$1 Gyr) populations, much more than the ETGs with 
 mid-IR excess (by a factor of seven). 
This amount of young component makes the E+As not in the red sequence of the 
 optical CMD.   
We note that the light fraction of $\sim$1 Gyr population for 
 the ETGs with mid-IR/near-UV excess reaches up to $\sim$20\%, 
 which explains why we could easily identify the ETGs with 
 recent star formation in the mid-IR/near-UV color space. 
Actually, the two-component SSP analysis of ETGs (Ko et al. 2013) showed
 that only $\sim$5\% of 1 Gyr population can make 
 significant changes in NUV$-r$ and [3.4]$-$[12] colors, thus we can easily 
 distinguish them from the ETGs without young populations.

\begin{figure}[!h]
\centering
\includegraphics[width=15cm]{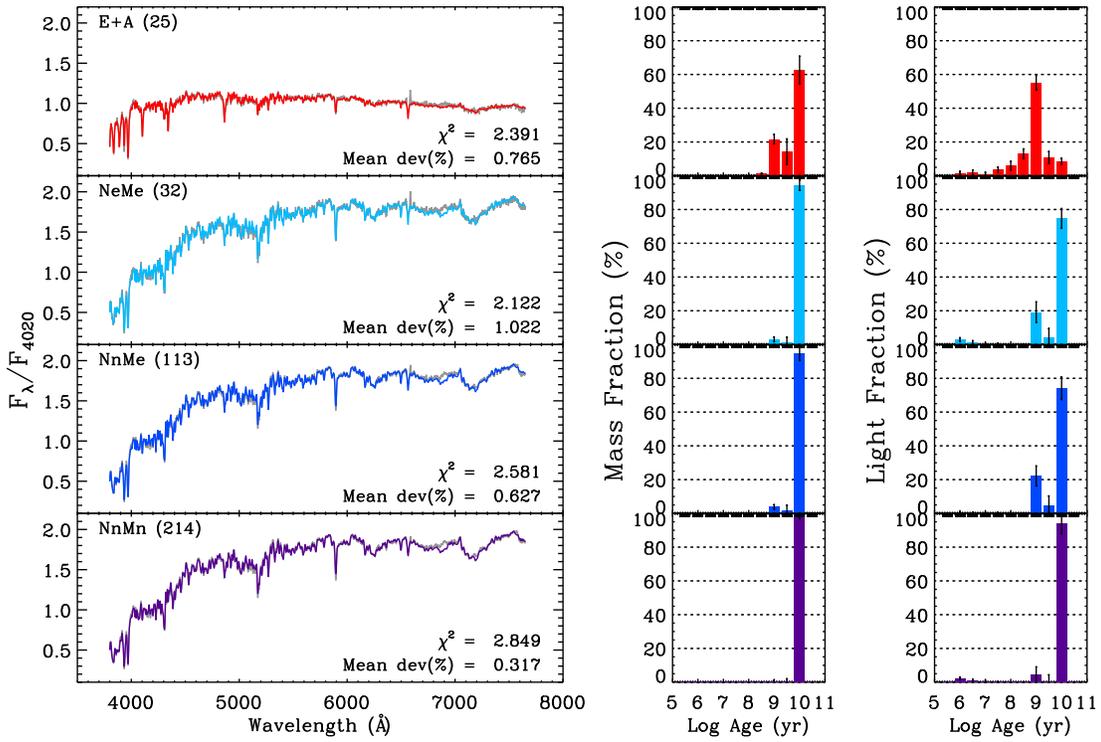}
\caption{STARLIGHT fit to the stacked spectra of the three 
         subclasses of ETGs and E+As (\textit{Left}).
         The gray and color-coded lines represent the stacked spectra 
         and the best-fit models, respectively.
         Two lines are not clearly visible because of overlap.         
         The fractions of stellar populations with 
         different ages in the total stellar mass (\textit{Middle}) and 
         in the total stellar light (\textit{Right}).
         Error bars indicate 1$\sigma$ distributions from the 100 fits 
         with different seeds.
         }         
\end{figure}

To examine the dependence of the fit on the choice of spectral templates, 
 we fit to the spectra with various combinations of age/metallicity 
 distributions (i.e., different star formation history), stellar population
 models (e.g., BC03 and Charlot \& Bruzual, in preparation, hereafter
 CB07\footnote{CB07 includes a new prescription of Marigo \& Girardi (2007) 
 for the thermally pulsating AGB evolution.}), and IMFs (e.g., Chabrier 2003 
 and Salpeter 1955). 
Figure 6 shows the resulting mass fractions from the fits with various age 
 and metallicity distributions of the templates. 
The plot shows that the mass fraction of young and intermediate-age stars 
 does not change much with different combinations of age and metallicity 
 distributions. 
Only when we use the templates with two metallicities without 
 the sub-solar one (see 2nd column in Figure 6), the mass fraction of 
 young and intermediate stars is about twice higher than the other cases. 

Figure 7 also shows the mass fractions of young and intermediate-age stars 
 from the fits, but with different stellar population models (BC03 or CB07)
 and IMFs (Chabrier or Salpeter). 
We fix the age and metallicity distributions of the templates to 
 the ones in Figure 5. 
The plot again shows that the mass fractions of young and
 intermediate-age stars do not differ much; 
 they are in the range $\sim$4$-$11\%.

\begin{figure}[!h]
\centering
\includegraphics[width=15cm]{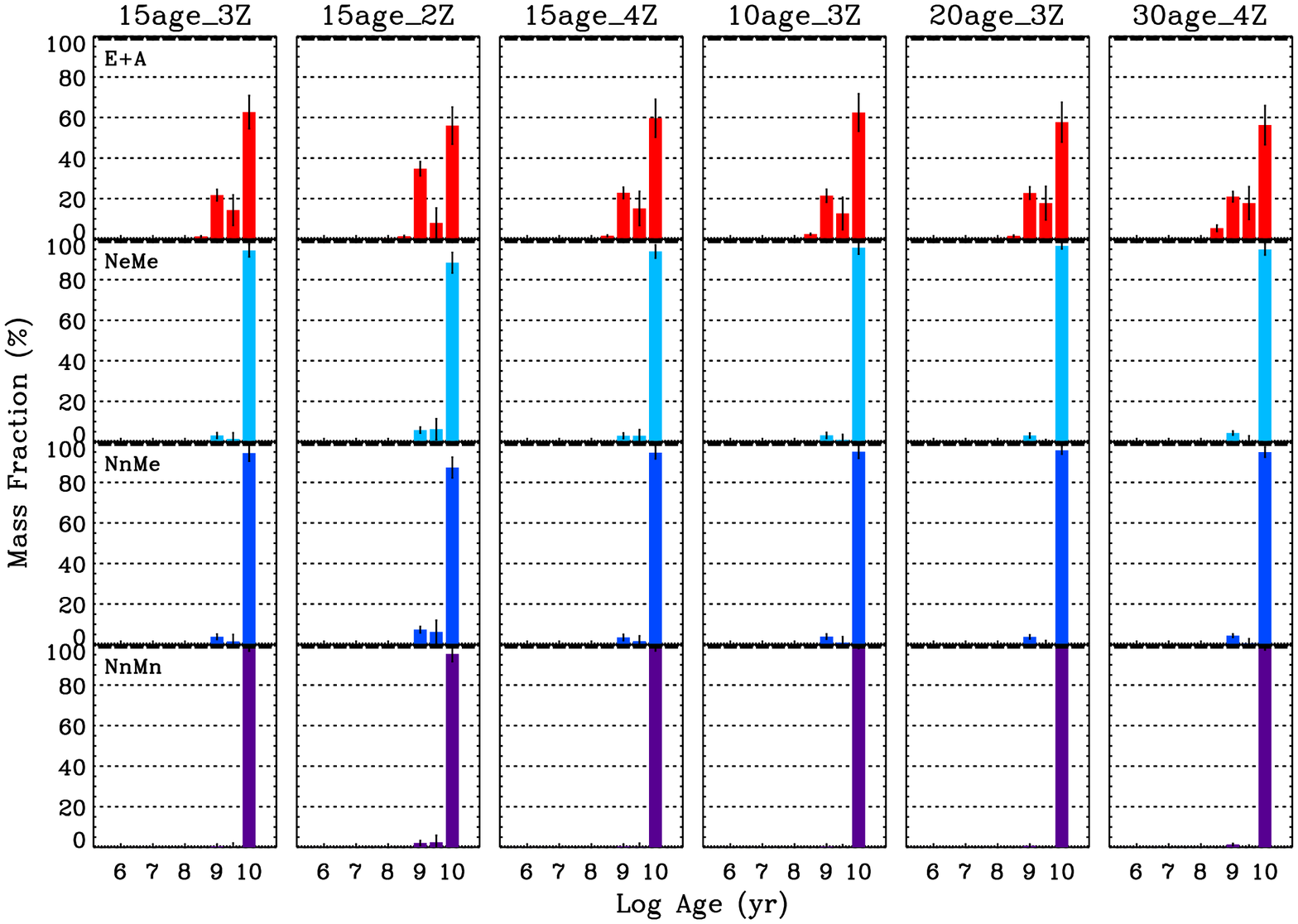}
\caption{The fractions of stellar populations with different ages in the 
		 total stellar mass for E+As (top panel) and three subclasses 
		 (bottom panels). Each column indicates the result from the fit 
		 with templates of different age and metallicity distributions. 
		 The title in each column indicates which template we use for the 
		 fit; the number in age means the number of age bins (i.e., 10, 15, 
		 20, 30) selected logarithmically from 1 Myr and 13 Gyr. The number 
		 in metallicity means the number of metallicity bins: 2 (Z=0.02, 
		 0.05), 3 (Z=0.004, 0.02, 0.05), and 4 (Z=0.004, 0.008, 0.02, 0.05).
         }         
\end{figure}

\begin{figure}[!h]
\centering
\includegraphics[width=15cm]{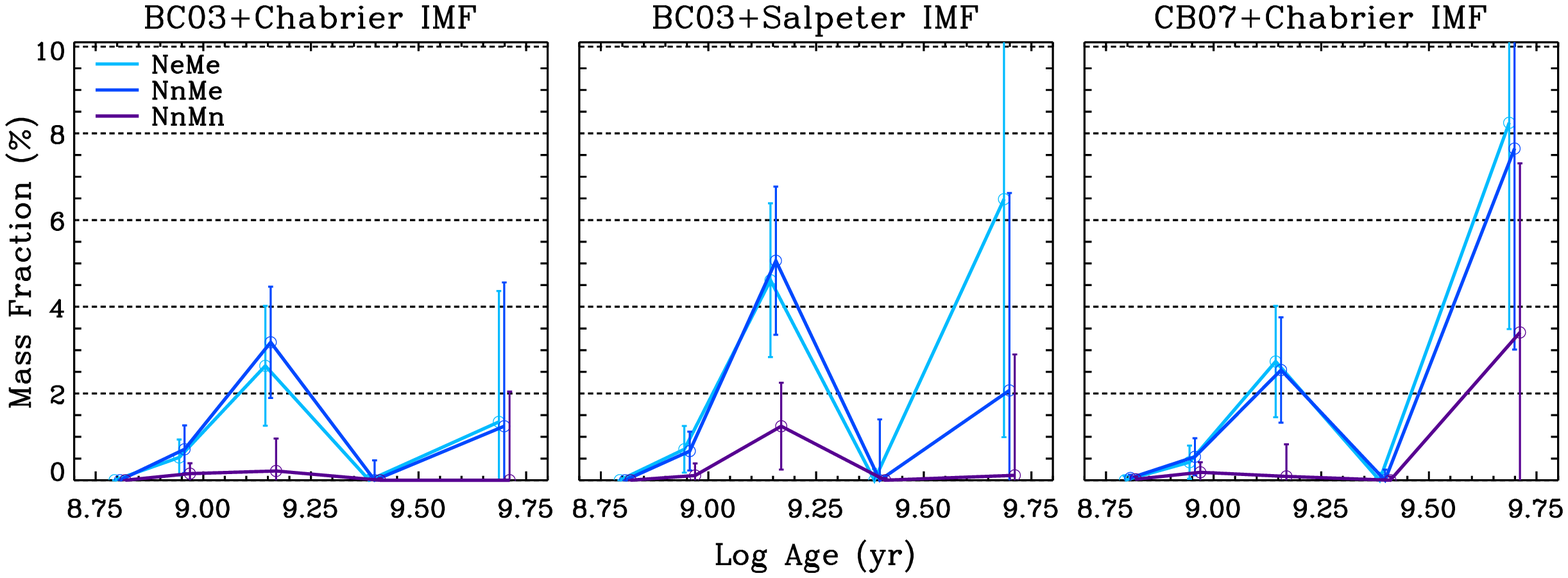}
\caption{Same as middle panels of Figure 5, but for
         young and intermediate-age populations of the 
         three subclasses of ETGs in three different combinations 
         of models and IMFs at the same age and metallicity 
         distributions as in Figure 5.
         }         
\end{figure}

Although the optical spectra of the ETGs in our sample do not show any 
 features of star-forming galaxies, there could still be a contribution to 
 the mid-IR excess emissions from deeply embedded sources (e.g., small,
 dense, central disks). 
However, among the 145 ETGs with mid-IR excess emission in our sample, 
 122 galaxies (84\%) are not detected at 22 $\mu$m 
 (i.e., signal-to-noise ratio at 22 $\mu$m $<$ 3); the remaining 23 ETGs 
 detected at 22 $\mu$m have small 22 $\mu$m flux densities (see Figure 7 
 of Ko et al. 2013). 
This suggests that most ETGs with mid-IR excess emission are not dusty, 
 star-forming systems. 
Instead, their mid-IR excess emission results probably from the existence 
 of young/intermediate age stars.

Interestingly, the fit to the stacked spectra shows that the stellar 
 populations of ETGs with both mid-IR and near-UV excess (NeMe) do not 
 significantly differ from those with only mid-IR excess (NnMe). 
The difference in intrinsic reddening between NeMe and NnMe is also 
 negligible (mean A$_{V}$ values derived from the 100 fits are 
 0.190 $\pm$ 0.017 and 0.192 $\pm$ 0.019, respectively), 
 suggesting that NnMe is not a simply dust-reddened NeMe.

There could be several explanations why NeMe galaxies have stronger NUV 
 emission than NnMe galaxies even though the stellar populations from 
 the spectra show no significant difference between the two. 
First, if recent star formation occurs $\gtrsim$1 Gyr ago in the
 central regions of galaxies (within the SDSS fiber 3$''$, 
 corresponding 2 kpc in radius at $z$ = 0.07) and 
 the dust attenuation effects are negligible, 
 the existence of extended UV structures (e.g., Salim et al. 2009) 
 in the outer regions could make the overall NUV$-r$ color 
 bluer (assuming an inside-out behavior of recent star formation).
In this case, galaxies with strong UV emission in the outer 
 regions could be detected as NeMe galaxies, but those with weak UV emission 
 in the outer regions would be classified as NnMe galaxies 
 (e.g., partial recent star formation; Jeong et al. 2009). 
However, the spectral features of NeMe and NnMe galaxies could be similar 
 because the optical spectra probe only the central regions where recent 
 star formation occurs around 1 Gyr ago in both populations.
The reason why NeMe and NnMe galaxies have different UV emissions in 
 the outer regions could be because NnMe galaxies are more evolved 
 systems than NeMe galaxies. 
In other words, NeMe galaxies have bluer NUV$-r$ colors because of recent 
 star formation occurred within $\sim$1 Gyr in the outer regions; 
 NnMe galaxies have relatively red NUV$-r$ colors because near-UV excess 
 fades with time (i.e., more than $\sim$1 Gyr after star formation stops), 
 but they still show mid-IR excess because of intermediate-age stars.
 
Second, if the dust attenuation effects are not negligible
 (particularly for NnMe) or/and dust geometry is different between the 
 two, the different NUV$-r$ colors can result from localized dust absorption.
In this case, NnMe galaxies probably contain dusty regions mixed with young 
 stars. 
Then their overall NUV$-r$ colors cannot be bluer because of dust attenuation
 even if the amount of young stars in NnMe galaxies is similar to that in 
 NeMe galaxies. 
This interpretation is supported by the fact that 70\% of 22$\mu$m-detected 
 ETGs with mid-IR excess are NnMe galaxies (i.e., 16 out of 23).






\section{CONCLUSIONS}
 
We study the optical spectral features of bright ($M_{r}$ $<$ $-$21.5)
 ETGs in the local universe that are on the optical red sequence but  
 have different amounts of near-UV and mid-IR excess emissions.
To quantify the amount of young and intermediate-age stars in those 
 ETGs, we perform a stellar population analysis using the SDSS stacked 
 spectra.
We find that the mid-IR colors ([3.4]$-$[12]) correlate well with
 the strength of $D_{n}$4000 and with the Balmer absorption line 
 H$\delta_{A}$ that are good tracers of star 
 formation activity over the past 1--2 Gyr.
These suggest that the mid-IR emission can be a useful diagnostic tool for 
 probing young and intermediate-age stars in ETGs. 
Actually there are various mechanisms that could contribute to the 
 mid-IR emission in galaxies (Draine \& Li 2007; da Cunha et al. 2008). 
This includes the SF driven dust continuum and PAH emissions 
 (Smith et al. 2007; Rieke et al. 2009), AGN heated hot dust continuum 
 (Netzer et al. 2007; Mullaney et al. 2011), and AGB dust emission 
 (Bressan et al. 1998; Piovan et al. 2003). 
Because we removed the galaxies with star formation and nuclear activity 
 from our sample (see Section 2.1), the contribution from the star formation
 and nuclear activity to the mid-IR emission is negligible in our sample.
This is also supported by that the majority (i.e., 122 out of 145) of the 
 ETGs with mid-IR excess emission are not detected at 22 $\mu$m (see Section
 3.3).

The fit to the stacked spectra with a spectral fitting code, STARLIGHT, 
 shows that the mass fraction of $\sim$1$-$5 Gyr stars in the ETGs with 
 mid-IR excess is $\sim$4$-$11\% and the light fraction 
 is up to $\sim20\%$.
It should be noted that the mass fraction of $\sim$1$-$5 Gyr stars 
 could be smaller than the current estimates (i.e., 4$-$11\%) if the 
 metallicity of ETGs is very low (e.g., Z$<$0.004). 
It is because the contribution of intermediate-age (i.e., 1$-$5 Gyr) AGB 
 stars to the mid-IR emission becomes small in a very low metallicity tail 
 (e.g., Villaume et al. 2015). 
In such cases, the mid-IR emission may not be a useful tool for probing 
 recent star formation. 
However, this is not a general case for nearby, massive ETGs including our
 sample that typically have metallicities close to solar value 
 (Gallazzi et al. 2006).
These results support the idea that the ETGs with mid-IR excess emission have 
 experienced star formation within the last 1$-$5 Gyr.

\begin{acknowledgements}

We thank the anonymous referee for useful comments that greatly improved this 
paper. J.K. and J.C.L. are the members of Dedicated Researchers for 
Extragalactic AstronoMy (DREAM) in Korea Astronomy and Space Science 
Institute (KASI).

This publication makes use of data products from the Wide-field Infrared Survey Explorer, which is a joint project of the University of California, Los Angeles, and the Jet Propulsion Laboratory/California Institute of Technology, funded by the National Aeronautics and Space Administration.
Funding for SDSS-III has been provided by the Alfred P. Sloan Foundation, the Participating Institutions, the National Science Foundation, and the U.S. Department of Energy Office of Science. The SDSS-III web site is http://www.sdss3.org/.
SDSS-III is managed by the Astrophysical Research Consortium for the Participating Institutions of the SDSS-III Collaboration including the University of Arizona, the Brazilian Participation Group, Brookhaven National Laboratory, Carnegie Mellon University, University of Florida, the French Participation Group, the German Participation Group, Harvard University, the Instituto de Astrofisica de Canarias, the Michigan State/Notre Dame/JINA Participation Group, Johns Hopkins University, Lawrence Berkeley National Laboratory, Max Planck Institute for Astrophysics, Max Planck Institute for Extraterrestrial Physics, New Mexico State University, New York University, Ohio State University, Pennsylvania State University, University of Portsmouth, Princeton University, the Spanish Participation Group, University of Tokyo, University of Utah, Vanderbilt University, University of Virginia, University of Washington, and Yale University.

\end{acknowledgements}

\bibliographystyle{apj} 

\clearpage

\end{document}